\documentclass[letter,twocolumn]{jpsj3}
\usepackage{txfonts}
\usepackage{url}
\usepackage{xcolor}

\usepackage[whole]{bxcjkjatype}
\usepackage{bm}
\usepackage{amsfonts}
\usepackage{physics}
\usepackage{float}
\usepackage{here}
\usepackage{comment}
\usepackage[ruled]{algorithm2e}
\usepackage{amssymb}
\usepackage{siunitx}
\usepackage{upgreek}

\usepackage{color}

\title{Utilizing intermediate states in quantum annealing for multi-objective optimization}

\author{Keita Takahashi$^{1}$ and Shu Tanaka$^{1,2,3,4,5}$\thanks{shu.tanaka@keio.jp}}
\inst{$^1$Graduate School of Science and Technology, Keio University, Kanagawa
223-8522, Japan \\
$^2$Department of Applied Physics and Physico-Informatics, Keio University, Kanagawa 223-8522, Japan \\
$^3$Keio University Sustainable Quantum Artificial Intelligence Center (KSQAIC), Keio University, Tokyo 108-8345,
Japan \\
$^4$Human Biology-Microbiome-Quantum Research Center (WPI-Bio2Q), Keio University, Tokyo 108-8345, Japan \\
$^5$Green Computing System Research Organization, Waseda University, Shinjuku, Tokyo 162-0042, Japan
}

\abst{
We investigate obtaining intermediate quantum states during the quantum annealing process to address the limitation of the linear weighted sum method in multi-objective optimization, which inherently fails to reach non-convex regions of the Pareto front. We validate this approach through physical experiments utilizing quench-based readout and numerical simulations assuming ideal mid-anneal measurements. Both methods consistently demonstrate a clear trade-off where  earlier timing enhances diversity of the solutions, whereas later timing ensures convergence to non-dominated solutions. Notably, a practical compromise timing balances both metrics. The qualitative agreement between practical quench and ideal simulation indicates the potential of accessing the intermediate states for comprehensive Pareto front exploration.
}

\kword{quantum annealing, mid-anneal measurement, quench, multi-objective optimization}

\begin{document}
\maketitle


\textit{Introduction---} Quantum annealing (QA) was proposed as a powerful metaheuristic method for solving combinatorial optimization problems by searching for the ground state of the Ising model~\cite{kadowaki1998quantum,tanaka2017quantum,tanahashi2019application,chakrabarti2023quantum}, and its applications have been widely explored since its hardware implementation~\cite{Johnson2011quantum}.
In recent years, the application of QA to multi-objective optimization (MOO) problems, which involve simultaneously optimizing conflicting objective functions, has attracted significant attention~\cite{King2025multiobjective, Aguilera2024multi}.
When applying QA to MOO, the most common approach is the linear weighted sum method~\cite{King2025multiobjective}. Since this approach reduces the problem to searching for the ground state of a composite Ising Hamiltonian, it is highly compatible with QA, which is inherently a single-objective optimization solver.
Consequently, many previous studies have employed this scalarization method and explored the set of Pareto optimal solutions by sweeping the weights.

However, the linear weighted sum method has a fundamental limitation: it cannot access solutions in the non-convex regions of the Pareto front.
In multi-objective optimization, the linear weighted sum method can identify only Pareto optimal solutions that are supported, i.e., those lying on the convex hull of the Pareto front in the objective space. 
Pareto optimal solutions in non-convex regions, known as unsupported Pareto optima, cannot correspond to the global minimum of the scalarized objective function for any choice of weights. 
Consequently, they never become the ground state of the composite Hamiltonian.
This implies that in the idealized setting where QA is assumed to perform an adiabatic ground-state search, such solutions are theoretically unreachable, severely limiting the diversity of the solution set obtained by conventional approaches.

To address this limitation, this study focuses on the intermediate quantum states during QA.
Herein, we refer to the intermediate quantum states as mid-anneal states.
During QA, the system evolves from a state dominated by quantum fluctuations to a classical state.
Crucially, at intermediate stages, the system exists in a superposition involving not only the ground state but also low-energy excited states of the problem Hamiltonian.
Therefore, performing measurements during the QA process, a method we refer to as mid-anneal measurement (MAM) in this study~\cite{Takahashi2025quantitative}, is expected to allow for the sampling of such low-energy excited states.
We hypothesize that Pareto optimal solutions in non-convex regions, which cannot correspond to the ground states in the linear weighted sum method, are encoded in the low-energy excited states of the problem Hamiltonian.
Based on this hypothesis, QA with mid-anneal measurement (MAM-QA) enables the sampling of diverse Pareto optimal solutions that are typically missed by conventional adiabatic ground state searches.
Therefore, this study aims to demonstrate that MAM-QA enhances solution diversity in MOO.
Specifically, we validate the effectiveness of the proposed method through two complementary approaches: experiments on a physical quantum annealer employing quench-based readout as a practical proxy, and numerical simulations of closed quantum systems assuming ideal MAM.


\textit{Quantum Annealing---} Quantum Annealing (QA) is a method for solving combinatorial optimization problems by searching for the ground state of an Ising model.
The cost function of the target problem is represented by the following problem Hamiltonian $H_{\mathrm{c}}$:
\begin{equation}
    H_{\mathrm{c}} = -\sum_{1 \leq i < j \leq N} J_{i,j} \sigma_{i}^{z} \sigma_{j}^{z} - \sum_{i=1}^N h_{i} \sigma_{i}^{z},
    \label{eq: Hc}
\end{equation}
where $N$ is the number of spins, $\sigma_{i}^{z}$ is the Pauli $z$ operator acting on the $i$-th spin, and $J_{i,j}$ and $h_{i}$ represent the interaction coefficients and longitudinal magnetic fields, respectively.
In QA, a transverse-field Hamiltonian $H_{\mathrm{q}} = -\sum_{i=1}^N \sigma_{i}^{x}$ introduces quantum fluctuations, and the total system is described by the following time-dependent Hamiltonian $H(s)$:
\begin{equation}
    H(s) = A(s) H_{\mathrm{q}} + B(s) H_{\mathrm{c}}, \quad s\in[0, 1],
    \label{eq: H(s)}
\end{equation}
where $s$ is the normalized annealing time, and $A(s)$ and $B(s)$ are the annealing schedules.
At the initial time $s=0$, we set $A(0) \gg B(0)$ and start from the ground state of $H_{\mathrm{q}}$, which is an equal superposition of all computational basis states.
According to the adiabatic theorem, if the system evolves slowly enough such that $A(1) \ll B(1)$, the ground state of $H_{\mathrm{c}}$, corresponding to the solution of the optimization problem, is obtained at the end of the annealing process~\cite{Kato1950adiabatic}.


\textit{Problem Setting---} In this study, we adopt the multi-objective minimization of Ising models with conflicting interactions as a benchmark.
Each objective of this problem is defined as:
\begin{equation}
    H_{\mathrm{obj}, m} = - \sum_{(i,j) \in E} w_{m, i, j} \sigma_i^z \sigma_j^z \quad (m=1, \dots, M),
\end{equation}
where $w_{m, i, j}$ represents the weight of interaction $(i, j)$ for the $m$-th objective function.
In the standard linear weighted sum method in QA, the problem Hamiltonian $H_\mathrm{c}$ is constructed using weight coefficients $\Omega_m \ge 0$ as:
\begin{equation}
    H_\mathrm{c} = \sum_{m=1}^{M} \Omega_m H_{\mathrm{obj}, m}, \quad \text{where } \sum_{m=1}^{M} \Omega_m = 1.
\end{equation}
To induce clear trade-offs between objective functions, we focus on the $M=2$ case and configure the interaction weights to create a conflict. Specifically, $w_{1, i, j}$ and $w_{2, i, j}$ are drawn uniformly from the intervals $[-1, 0]$ and $[0, 1]$, respectively. In this study, we ensure that the problem graph is connected.


\textit{Mid-anneal Measurement QA (MAM-QA)---} Conventional QA aims to obtain the ground state of $H_\mathrm{c}$ at the final time $s=1$ by gradually decreasing the coefficient $A(s)$ of the transverse field term $H_{\mathrm{q}}$ and increasing the coefficient $B(s)$ of the problem Hamiltonian $H_\mathrm{c}$.
In contrast, MAM-QA exploits quantum superposition at intermediate stages of the QA process to sample solutions corresponding to low-energy excited states of the problem Hamiltonian~\cite{Takahashi2025quantitative}.
When using the linear weighted sum method in MOO, solutions located in the non-convex regions of the Pareto front cannot become the ground state of $H_\mathrm{c}$ under any combination of weights $\Omega_m$, making them theoretically inaccessible.
To address this issue, we employ MAM-QA.
Based on the hypothesis that non-convex Pareto optimal solutions are encoded in the low-energy excited states of $H_\mathrm{c}$, we attempt to access these solutions by performing measurements at intermediate annealing times $s < 1$.

\textit{Experimental Settings---} To verify the effectiveness of the proposed method, we conducted experiments using a physical quantum annealer and numerical simulations.
We performed measurements at multiple points in the normalized time range $s \in [0, 1]$. Measurements at $s=0$ correspond to an equal-weight superposition dominated by quantum fluctuations, those at $0 < s < 1$ correspond to MAM-QA, and those at $s=1$ correspond to standard QA (normal QA).
For each measurement, we discretized $\Omega_1$ into 101 equally spaced points between 0 and 1, collecting 1000 samples at each point.

\begin{figure}[t]
    \centering
    \includegraphics[width=1\linewidth]{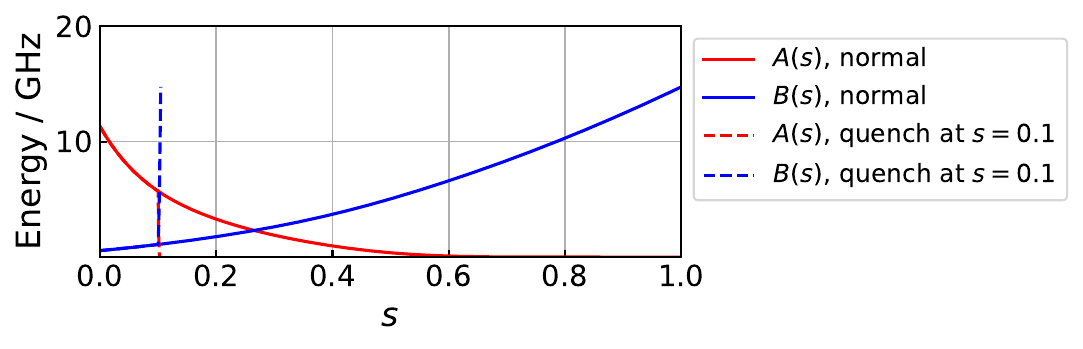}
    \caption{Annealing schedules for normal QA and QA with quench-based readout at $s=0.1$.}
    \label{fig: schedule}
\end{figure}

As the physical quantum annealer, we utilized the \texttt{Advantage2\_system1.10}. 
Because current devices lack a direct mid-anneal projective measurement capability, the intermediate quantum state was probed indirectly via a quench operation, in which the annealing parameters $A(s)$ and $B(s)$ are abruptly changed to their final values $A(1)$ and $B(1)$ at the target time $s$.
In the following, we refer to this procedure as a quench-based readout. 
Quench-based readout has been used as an effective probe of intermediate states during QA in previous studies.
King et al.~\cite{king2018observation} used quench dynamics to probe intermediate states and observed topological phenomena. 
While the quench is not strictly instantaneous and may induce some non-equilibrium effects, they reported good agreement between experimentally measured observables obtained via quench-based readout and theoretical predictions based on quantum Monte Carlo simulations. 
Motivated by these results, we assume that the quench-based readout provides an effective probe of the intermediate states.
Figure~\ref{fig: schedule} shows examples of the annealing schedules for normal QA and QA with quench-based readout used in the hardware experiments.
For the problem graph, we employed a single instance of Zephyr graph ($N=4589$) native to the machine's topology, with interaction weights ($w_{1, i, j}$ and $w_{2, i, j}$) generated as described in the Problem Setting. This graph size was selected to clearly distinguish the behaviors of normal QA and QA with quench-based readout, given the finite time resolution of the quench operation relative to the annealing dynamics. The annealing time was fixed at $100\,\upmu\mathrm{s}$. Thus, the normalized time $s=1$ corresponds to $100\,\upmu\mathrm{s}$. For instance, in the case of QA with quench-based readout at $s=0.1$, the quench operation is executed at $10\,\upmu\mathrm{s}$.

Furthermore, to exclude hardware noise and obtain theoretical insights, we performed sampling simulations of an isolated system using QuTiP~\cite{Johansson2012qutip, Johansson2013qutip}.
To model the ideal adiabatic limit, we calculated the instantaneous ground state $\ket{\psi_\mathrm{GS}(s)}$ of the total Hamiltonian $H(s) = A(s)H_\mathrm{q} + B(s)H_\mathrm{c}$ at each time $s$, which represents the quantum state followed during adiabatic evolution at the corresponding annealing time.
The coefficients $A(s)$ and $B(s)$ were set based on the physical hardware schedule shown in Fig.~\ref{fig: schedule}.
Samples were generated according to the probability distribution $|\langle z | \psi_\mathrm{GS}(s) \rangle|^2$, which corresponds to an ideal projective measurement of the instantaneous quantum state in the computational basis $\ket{z}$ within the closed-system model.
For the problem graph, considering computational costs, we adopted small-scale fully connected undirected graphs ($N=6, 8, 10$) and generated 10 instances.

\textit{Evaluation Metrics---} To quantitatively evaluate the quality of the solution sets in MOO, we employ the following three metrics.
For each measurement timing, the solution set used for evaluation is constructed by aggregating all sampled solutions over the entire range of weight coefficients $\{\Omega_m\}$.
\begin{itemize}
    \item Hypervolume (HV): This metric represents the volume of the objective function space dominated by the solution set, capturing both convergence to the Pareto front and solution diversity. Higher values indicate better performance.
    In this study, HV is computed with respect to a common reference point defined by the worst (component-wise maximum) objective values observed over all measurement timings and weight coefficients. HV is then normalized by the value at $s = 0$, which corresponds to an equal-weight superposition in the ideal closed-system model, evaluated using the same reference point. We calculate HV following Ref.~\cite{Zitzler1999evolutionary} after removing duplicate solutions.
    \item Spacing (SP): This metric quantifies the uniformity of the solution distribution in the objective space. Lower values indicate better uniformity, signifying that solutions are distributed at nearly equal intervals. We normalize the values relative to the SP obtained from normal QA (the value at $s=1$), following the method in Ref.~\cite{Schott1995fault} after removing duplicate solutions.
    \item Ratio of non-dominated individuals (RNI): This metric is defined as the ratio of non-dominated solutions to the total number of sampled solutions, counting all occurrences including duplicates. Here, the reference set of non-dominated solutions is identified from the union of all solution sets obtained across all measurement timings and weight coefficients. RNI primarily serves as an indicator of convergence, where higher values indicate better convergence. The calculation is based on Ref.~\cite{Tan2001incrementing}.
\end{itemize}

\textit{Results---} We present the results of experiments using the physical quantum annealer and numerical simulations.
We evaluate the impact of the MAM timing $s$ on the diversity and convergence of solutions by comparing the solution sets obtained by MAM-QA (implemented via quench-based readout on the physical quantum annealer) and normal QA.

\begin{figure}[t]
    \centering
    \includegraphics[width=1\linewidth]{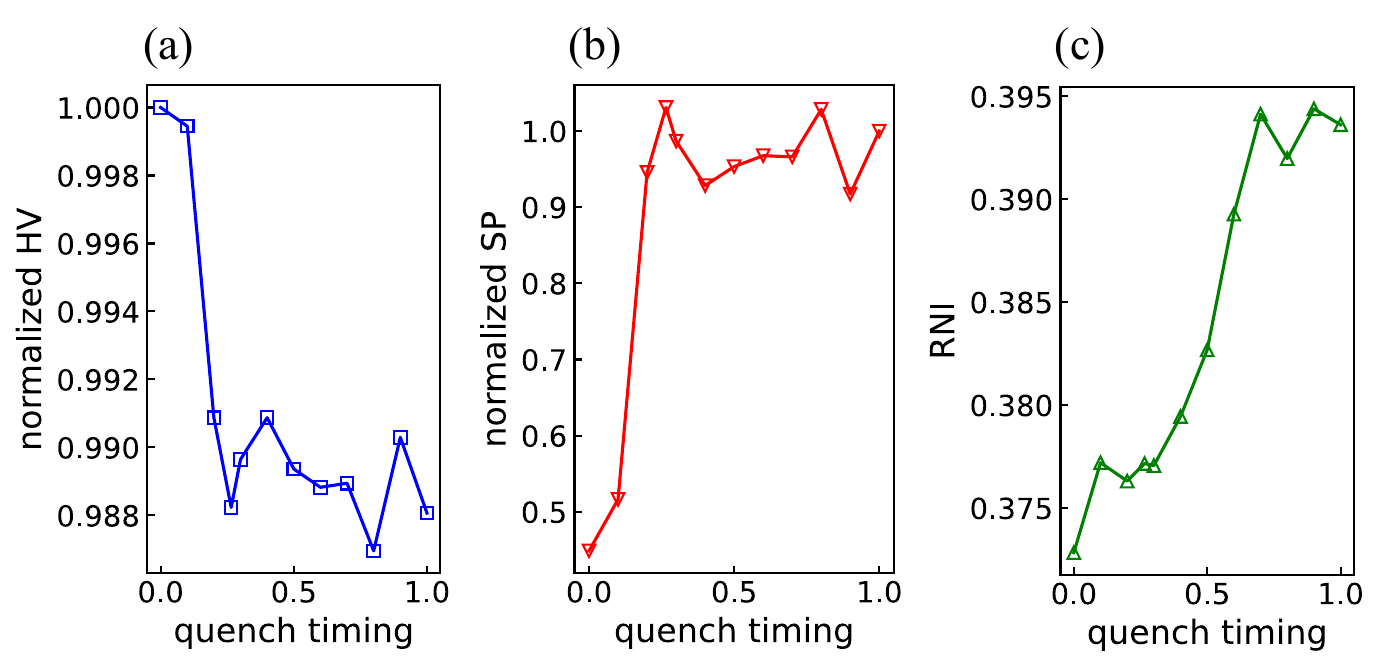}
    \caption{Dependence of evaluation metrics on the quench-based readout timing $s$ in physical quantum annealer experiments: (a) normalized HV, (b) normalized SP, and (c) RNI.}
    \label{fig: dwave metrics}
\end{figure}

First, we discuss the results obtained from the physical quantum annealer.
To quantitatively evaluate the quality of the solution sets, Fig.~\ref{fig: dwave metrics} illustrates the dependence of each evaluation metric on the quench timing $s$.
As shown in Figs.~\ref{fig: dwave metrics}(a) and (b), HV and SP exhibit superior values when the quench is performed at early timings.
Conversely, Fig.~\ref{fig: dwave metrics}(c) indicates that RNI reaches higher values during the latter half of annealing schedule or in normal QA ($s \to 1$).
Notably, HV and SP exhibit a similar trend in this experimental setting. This is because the solution set is distributed along a single quasi-one-dimensional curve, meaning that variations in HV are primarily determined by the spread of the solutions.
These quantitative results reveal a clear trade-off between solution ``diversity'' and ``convergence'' with respect to the quench timing.

\begin{figure}[t]
    \centering
    \includegraphics[width=1\linewidth]{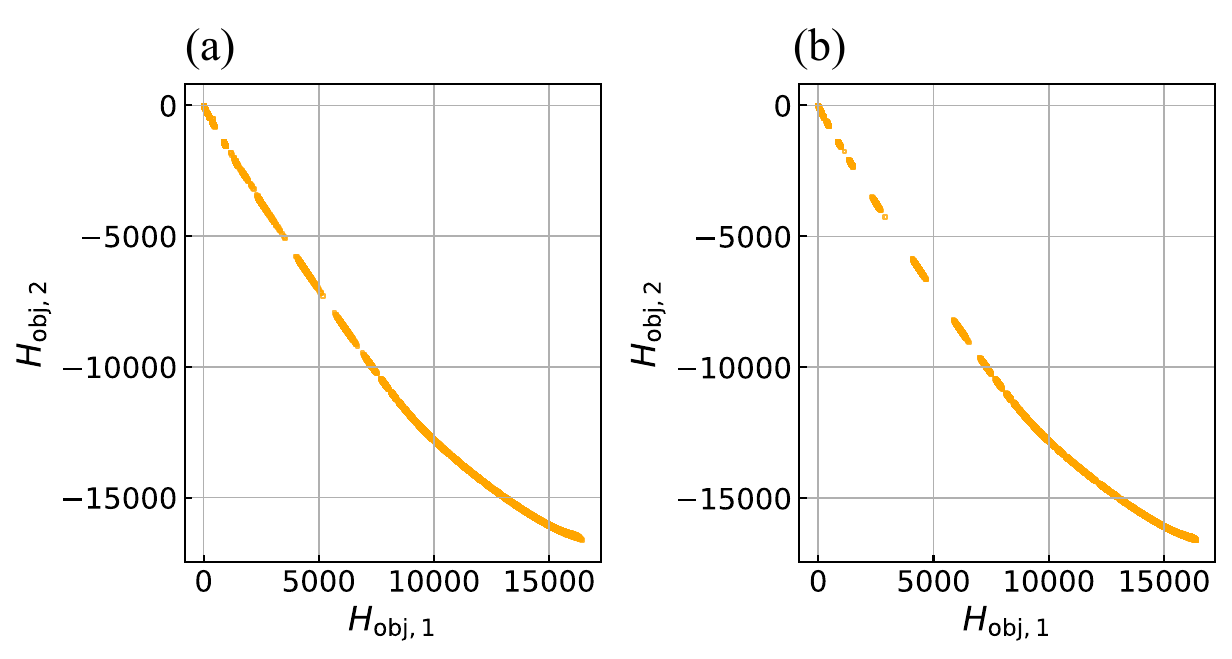}
    \caption{Sampling results using a physical quantum annealer. (a) QA with quench-based readout at $s=0.1$, (b) normal QA.}
    \label{fig: dwave scatter}
\end{figure}

Figure~\ref{fig: dwave scatter} displays the distribution of solutions in the objective function space obtained using the physical quantum annealer.
We observe that quench-based readout at early timings (e.g., $s=0.1$ in Fig.~\ref{fig: dwave scatter}(a)) yields a solution distribution covering a wider region in the objective space than normal QA ($s=1$) shown in Fig.~\ref{fig: dwave scatter}(b). 
This qualitatively confirms that probing intermediate states of quantum annealing via quench enhances solution diversity compared to normal QA.

\begin{figure}[t]
    \centering
    \includegraphics[width=1\linewidth]{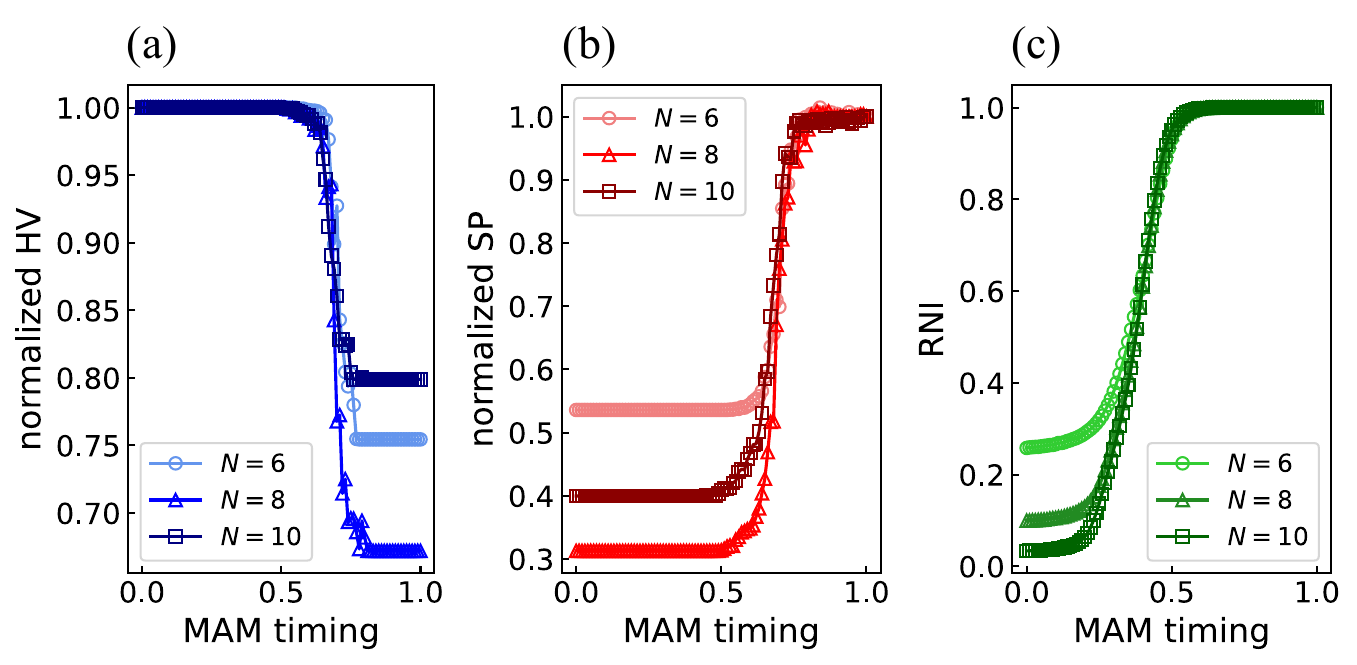}
    \caption{Dependence of evaluation metrics on mid-anneal measurement timing $s$ in closed-system simulations. Solid lines show average over 10 instances. (a) normalized HV, (b) normalized SP, and (c) RNI.}
    \label{fig: sim metrics}
\end{figure}

Next, to verify whether the trade-off observed in the hardware experiments originates from the fundamental characteristics of QA, we performed sampling simulations of a closed system in the adiabatic limit, assuming ideal projective measurements (MAM).
Figure~\ref{fig: sim metrics} shows the dependence of each evaluation metric on the mid-anneal measurement timing in the simulation.
As shown in Fig.~\ref{fig: sim metrics}, the simulation results confirm the trade-off between diversity and convergence, similar to the trend observed in the hardware experiments.
Specifically, HV and SP exhibit superior performance at earlier measurement timings (Figs.~\ref{fig: sim metrics}(a) and (b)), whereas RNI improves towards the end of annealing ($s=1$) (Fig.~\ref{fig: sim metrics}(c)).
Furthermore, while a trade-off between diversity and convergence is observed overall, it is noteworthy that there exists a compromise MAM timing where all metrics exhibit balanced and favorable values for multiple problem sizes $N$.
The existence of a practical compromise timing that reconciles diversity and convergence strongly supports the practical effectiveness of MAM-QA for MOO.

\begin{figure}[t]
    \centering
    \includegraphics[width=1\linewidth]{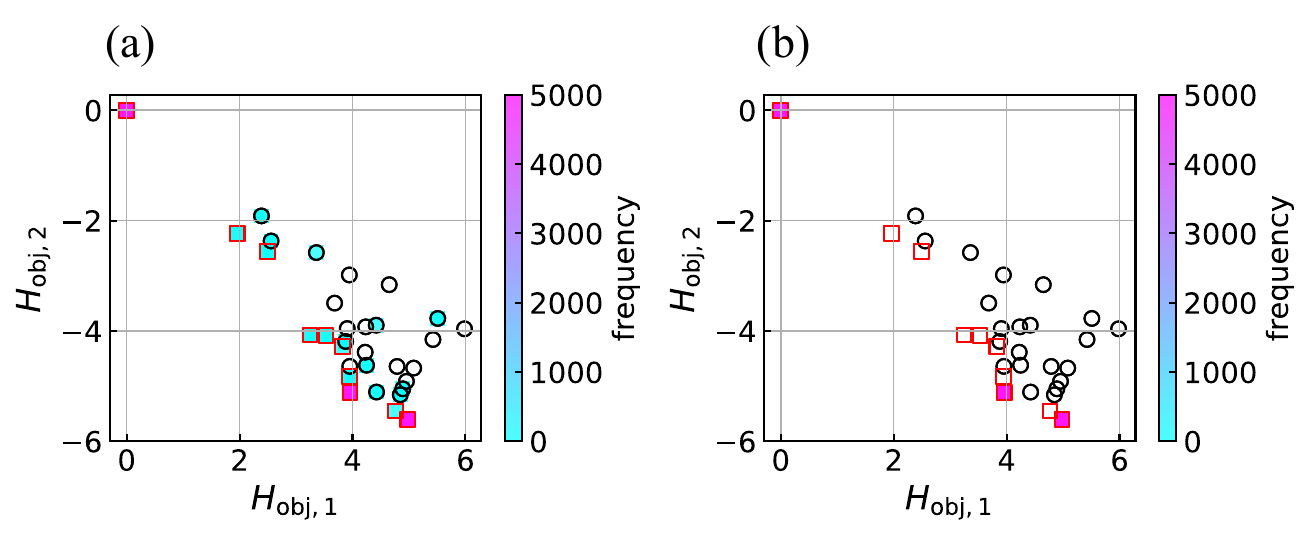}
    \caption{Simulation sampling results for a problem instance ($N=6$). Red squares indicate Pareto-optimal solutions identified by exhaustive enumeration of all $2^N$ configurations (minimization convention), while black circles denote non-Pareto solutions. The color bar represents the sampling frequency of each solution. (a) MAM-QA at $s=0.6$, (b) normal QA.}
    \label{fig: sim scatter}
\end{figure}

To visually verify the quality of the solution set at the practical compromise timing, we focus on MAM-QA at $s=0.6$ as a representative example. Fig.~\ref{fig: sim scatter} shows the distribution of solutions in the objective function space obtained from the simulation.
For this $N = 6$ instance, we exhaustively enumerated all $2^N$ spin configurations to identify the true Pareto-optimal solutions under the minimization convention (red squares in Fig.~\ref{fig: sim scatter}).
From Fig.~\ref{fig: sim scatter}, it is evident that MAM-QA successfully acquires solutions in the non-convex regions of the Pareto front, which were not obtained by normal QA.
A key direction for future research is to investigate the quantitative relationship between solutions in non-convex regions of the Pareto front and the energy levels of the corresponding excited states of the problem Hamiltonian.
Clarifying specifically which energy levels these solutions occupy will provide deeper insights into the working mechanism of MAM-QA.


\textit{Conclusion---} In this study, we investigated the strategy of obtaining intermediate quantum states in quantum annealing to address the inherent limitation of the conventional linear weighted sum method in multi-objective optimization (MOO), which is the inability to access solutions in non-convex regions of the Pareto front.
We validated this concept through two complementary approaches: physical experiments employing quench-based readout as a practical proxy for measurement, and numerical simulations assuming ideal mid-anneal measurements (MAM) in the adiabatic limit.

The experimental results using quench-based readout demonstrated a significant improvement in solution diversity at early annealing stages. Furthermore, the numerical simulations assuming ideal MAM confirmed that this approach enables access to solutions in the non-convex regions of the Pareto front, which are inaccessible to normal QA. 
Both the experiments and the simulations consistently elucidated a clear trade-off between ``diversity'' and ``convergence,'' governed by the measurement timing.
Specifically, measurements at early stages significantly improved solution diversity, whereas convergence to the non-dominated solutions was maximized under normal QA.
Importantly, our results identified the existence of a practical compromise timing, where both diversity and convergence metrics exhibit balanced and favorable values.
This suggests that accessing intermediate states can serve as an effective strategy for obtaining high-quality solutions for MOO.
Furthermore, for a more comprehensive and high-precision search, we propose a hybrid strategy that combines normal QA to ensure convergence with MAM-QA (implemented via quench-based readout or future direct readout technologies) to ensure diversity.
In future work, we aim to apply this approach to practical real-world optimization problems to further demonstrate its utility.

\section*{acknowledgement}
This work was partially supported by the Japan Society for the Promotion of Science (JSPS) KAKENHI (Grant Number JP23H05447), the Council for Science, Technology, and Innovation (CSTI) through the Cross-ministerial Strategic Innovation Promotion Program (SIP), ``Promoting the application of advanced quantum technology platforms to social issues'' (Funding agency: QST), Japan Science and Technology Agency (JST) (Grant Number JPMJPF2221). The computations in this work were partially performed using the facilities of the Supercomputer Center, the Institute for Solid State Physics, The University of Tokyo. S. Tanaka wishes to express their gratitude to the World Premier International Research Center Initiative (WPI), MEXT, Japan, for their support of the Human Biology-Microbiome-Quantum Research Center (Bio2Q).

\bibliographystyle{jpsj}
\bibliography{references}

\end{document}